\documentclass[lettersize,correspondense]{IEEEtran}

\usepackage{amsmath,amsfonts}
\usepackage{array}
\usepackage[caption=false,font=normalsize,labelfont=sf,textfont=sf]{subfig}
\usepackage{textcomp}
\usepackage{stfloats}
\usepackage{url}
\usepackage{verbatim}
\usepackage{graphicx}
\usepackage{epstopdf}
\usepackage{amsmath}
\usepackage{algorithm}
\usepackage{algorithmicx}
\usepackage{algpseudocode}
\usepackage{float}

\usepackage[utf8]{inputenc} 

\usepackage{booktabs} 

\usepackage{cite}

\usepackage{amssymb}

\usepackage[colorlinks=true,
            linkcolor=black, 
            anchorcolor=black, 
            citecolor=black, 
            urlcolor=blue
            ]{hyperref}

\def\BibTeX{{\rm B\kern-.05em{\sc i\kern-.025em b}\kern-.08em
    T\kern-.1667em\lower.7ex\hbox{E}\kern-.125emX}}
\usepackage{balance}
\begin{document}
\bibliographystyle{ieeetr}
\newcommand{\upcite}[1]{\textsuperscript{\cite{#1}}}

\title{LLM-RIMSA: Large Language Models driven Reconfigurable Intelligent Metasurface Antenna Systems}
\author{Yunsong Huang, Hui-Ming Wang \emph{Senior Member, IEEE}, Qingli Yan and Zhaowei Wang


\thanks{
	
	Yunsong Huang, Hui-Ming Wang, and Zhaowei Wang are with the School of Information and Communications Engineering, Xi'an Jiaotong University,
	Xi'an 710049, China (e-mail: song1102@stu.xjtu.edu.cn; xjbswhm@gmail.com; wangzwssd@stu.xjtu.edu.cn).}
\thanks{
	Qingli Yan is with the School of Computer Science \& Technology,  Xi'an University of Posts \& Telecommunications, Xi'an 710121, Shaanxi, China (e-mail: yql@xupt.edu.cn)}

}

\maketitle

\begin{abstract}
The evolution of 6G networks demands ultra-massive connectivity and intelligent radio environments, yet existing reconfigurable intelligent surface (RIS) technologies face critical limitations in hardware efficiency, dynamic control, and scalability. This paper introduces LLM-RIMSA, a transformative framework that integrates large language models (LLMs) with a novel reconfigurable intelligent metasurface antenna (RIMSA) architecture to address these challenges. Unlike conventional RIS designs, RIMSA employs parallel coaxial feeding and 2D metasurface integration, enabling each individual metamaterial element to independently adjust both its amplitude and phase. While traditional optimization and deep learning (DL) methods struggle with high-dimensional state spaces and prohibitive training costs for RIMSA control, LLM-RIMSA leverages pre-trained LLMs cross-modal reasoning and few-shot learning capabilities to dynamically optimize RIMSA configurations. 
Simulations demonstrate that LLM-RIMSA achieves state-of-the-art performance, outperforming conventional DL-based methods in sum rate while reducing training overhead. The proposed framework pave the way for LLM-driven intelligent radio environments.

\end{abstract}


\section{Introduction}
\IEEEPARstart{T}{he}
sixth-generation (6G) wireless networks aim to support ultra-massive connectivity for emerging applications like Internet of things (IoT) ecosystems, holographic communications, and tactile internet, which demand unprecedented spectral efficiency, sub-millisecond latency, and three-dimensional coverage \cite{chen2024big}. While traditional massive MIMO and millimeter-wave technologies enhance spectral efficiency, phased arrays face critical limitations in hardware cost, power consumption, and energy efficiency, particularly as antenna scales expand to extra-large MIMO (XL-MIMO). 
These challenges have spurred the development of reconfigurable intelligent surfaces (RIS) \cite{J2024Intelligent}, a disruptive hardware paradigm that dynamically manipulates electromagnetic wave propagation to enable low-cost, energy-efficient, and intelligent wireless environments \cite{T2023Holographic}. 

\subsection{Technological Evolution of RIMSA}
RIS technologies are broadly categorized into three types, each with unique challenges and application scenarios. The first type is reflective RIS, which passively reflects incident signals to enhance coverage without relying on radio frequency (RF) chains \cite{M2020Smart}. 
The programmable manipulation of reflected signal phase and amplitude enables precise beam steering through RIS. Emerging research has validated RIS transformative potential across key wireless communication metrics: channel capacity maximization \cite{Q2019Intelligent}-\cite{S2020Capacity}, energy consumption minimization \cite{C2019Reconfigurable}, and secure transmission via physical-layer encryption \cite{L2020Enhancing}.
However, this approach suffers from double-path fading. Active RIS variants address this issue by amplifying signals but introduce high power consumption and increased hardware complexity \cite{W2023RIS}. 

The second type is transmissive RIS, which directs signals through the metasurface to fill coverage gaps. Nevertheless, it incurs significant insertion loss, reducing aperture efficiency \cite{S2021Reconfigurable}. The third type is radiative RIS , such as dynamic metasurface antennas (DMA) and reconfigurable holographic surfaces (RHS), which integrate radiation and phase-shifting capabilities \cite{N2019Dynamic}. However, DMA waveguide-based architecture introduces frequency selectivity due to Lorentzian-constrained elements and propagation delays \cite{M2024Channel}, while RHS relies on holographic interference patterns that are limited to amplitude modulation  \cite{R2021Reconfigurable}. Furthermore, both approaches employ serial feeding, leading to uneven excitation and increased control complexity.

To address the limitations of existing RIS technologies, we propose the reconfigurable intelligent metasurface antenna (RIMSA), a novel radiative RIS architecture. At its core, RIMSA employs parallel coaxial feeding, a departure from traditional serial-fed designs. This innovative feeding mechanism excites all metasurface elements simultaneously through a power distribution network, effectively eliminating frequency selectivity and propagation delays that have long plagued systems like DMA \cite{T2021Channel}. 
A key feature of RIMSA is its incorporation of amplitude-phase joint modulation, which enables each individual metamaterial element to independently adjust both its amplitude and phase. Importantly, this independent control over amplitude and phase for every single unit is critical to achieving high-precision beamforming, enhancing signal directivity, and enabling efficient multi-user interference suppression. Such capabilities make RIMSA particularly well-suited for dense communication environments where fine-grained control and adaptability are essential \cite{A2014Channel}. 
Additionally, RIMSA adopts a compact 2D integration design, where the metasurface serves as the radiating aperture. By avoiding bulky waveguides, this design significantly reduces hardware complexity and physical footprint, supporting scalable deployment in compact and resource-constrained environments. 

\subsection{Limitations of Model and DL Driven RIS Control Methods}
Despite these advantages, scaling such systems to support thousands of reconfigurable elements remains hindered by unresolved control challenges \cite{H2024Modelbased}.
Traditional optimization-based RIS control methods \cite{J2022Robust} excel in model-driven frameworks via channel estimation-parameter optimization paradigms. Representative techniques like convex relaxation \cite{H2020Intelligent}, block coordinate descent \cite{Z2023Robust}, and alternating optimization \cite{H2022Robust} achieve near-theoretical-limit performance (e.g., sum-rate \cite{A2023Reconfigurable}) under precise channel state information (CSI), offering convergence guarantees and computational efficiency in low-dimensional scenarios \cite{X2022Joint}. However, they face critical limitations: First, explicit channel estimation incurs prohibitive pilot overhead for large-scale systems \cite{C2019Reconfigurable}; Second, non-convex constraints (e.g., unit modulus) amplify complexity, while static formulations lack adaptability to dynamic environments \cite{A2024Age}. These scalability and practicality bottlenecks hinder deployment in complex wireless networks \cite{R2022AI}.

Deep learning (DL) based approaches have emerged as innovative solutions for controlling RIS, demonstrating significant advantages in model-free optimization \cite{M2024comprehensive}. 
Among these DL methods, deep reinforcement learning (DRL) \cite{C2021Multi} and graph neural networks (GNNs) \cite{T2021Learning} stand out as representative techniques \cite{J2021Interplay}.
DRL method models RIS configuration through a markov decision process \cite{W2022Intelligent}, this approach dynamically adapts to complex radio environments without requiring sub-channel CSI, significantly enhancing system performance metrics like sum rate \cite{Y2025Pilot}. GNN architecture implicitly estimate channels, directly mapping received pilot signals to optimal beamforming and reflection coefficients \cite{Z2022Learning}. The permutation invariant/equivariant properties of GNNs effectively capture multi-user interference while incorporating heterogeneous data like user locations, achieving near-perfect CSI performance under limited pilot lengths \cite{B2025Distributed}. However, DRL suffers from dependency on extensive environmental interaction data and slow training convergence, with potential inefficiencies in exploring high-dimensional RIS phase spaces \cite{Z2025Malicious}. GNNs, despite their interpretability, face escalating computational complexity as user and reflection unit quantities increase, coupled with reliance on offline training data and limited adaptability to dynamic channel variations \cite{B2023Graph}. 


Both approaches must overcome critical deployment challenges: the complex high-dimensional control demands arising from thousands of RIMSA elements requiring precise amplitude phase optimization, and the need for real-time adaptation to dynamic environments characterized by time-varying channels and multi-user interference under constrained pilot resources. These persistent limitations highlight the necessity of further studies to improve system robustness and ensure scalable implementations for large-scale RIMSA deployments.

\subsection{LLM Driven RIMSA Control}
Large language models (LLMs) \cite{K2024Large} are finding diverse applications in wireless communications, enhancing various aspects of network performance and user experience \cite{L2024large}. They enable intuitive customer support, protocol optimization through network scenario simulation \cite{Du2024enabling}, and advanced network analytics for fault detection and automated reporting \cite{L2025LAMBO}. At the physical layer, LLMs optimize critical parameters like beamforming directions to improve signal coverage \cite{F2024Large}. LLM4CP demonstrates how fine-tuned GPT achieves high-precision channel prediction by aligning channel characteristics with LLM feature spaces \cite{Liu2024LLM4CP}.

LLMs bridge semantic environmental understanding with physical-layer beamforming, revolutionizing RIMSA control. Unlike model-driven approaches, LLMs infer latent relationships between dynamic channel states and optimal RIMSA configurations through context-aware reasoning \cite{H2024Large,J2024Large}. Pre-trained LLMs, fine-tuned on sparse experimental data, generalize beamforming policies across diverse deployment scenarios \cite{M2023Timellm}, enabling real-time, model-agnostic control and transforming passive hardware into self-optimizing cognitive systems \cite{Y2024TPLLM}. Compared to traditional methods, LLMs exhibit superior scalability and computational efficiency by processing heterogeneous data streams \cite{Liu2024LLM4CP}, while their continuous learning capabilities ensure robustness against time-varying channels and user mobility \cite{A2024dawn}. Additionally, LLMs exposure to diverse training corpora enhances pattern recognition, enabling deeper insights into propagation mechanisms through semantic-channel feature alignment \cite{jiang2024large}.

Recent advancements in LLMs have demonstrated their exceptional ability to model complex temporal and spatial relationships in cross-modal tasks.  
Different from prompt-based WirelessLLM proposed in \cite{J2024WirelessLLM}, which focuses on integrating multimodal data such as text, signals, and protocols, and enhancing domain knowledge through retrieval-augmented generation (RAG) and external tools, LLM-RIMSA emphasizes the design of specialized modules to adjust the model architecture to the input feature space of LLMs. Furthermore, LLM employs parameter freezing and fine-tuning strategies to retain the general knowledge of the pre-trained model while enhancing its generalization capabilities. As a result, compared to WirelessLLM, LLM-RIMSA demonstrates distinct advantages in achieving high prediction accuracy and strong generalization performance.
By leveraging the universal modeling capabilities of LLMs, we can bypass explicit channel estimation, and reduce dependency on exhaustive trial-and-error learning. 
This paper proposes LLM-RIMSA, the first framework to harness LLMs for real-time RIMSA configuration in wireless systems. Our key contributions are:

\begin{itemize}
	\item [$\bullet$]RIMSA architecture innovation: Introduces a novel RIMSA architecture with a parallel coaxial feeding network and 2D metasurface integration, detailing its structural design and signal modeling as both transmitting and receiving antennas.
	\item [$\bullet$]LLM-driven framework: Proposes the first LLM-RIMSA framework leveraging a GPT-2 backbone to dynamically optimize RIMSA configurations using pilot signals, eliminating explicit sub-channel CSI estimation. The framework bridges high-dimensional channel data and metasurface control through end-to-end optimization.
	\item [$\bullet$]Cross-modal data processing: Introduces a space domain preprocessing module and positional embedding layers to adapt LLMs for high-dimensional RIMSA parameter control while preserving spatiotemporal correlations in wireless channels.
	\item [$\bullet$]Performance validation: Simulations demonstrate LLM-RIMSA achieves higher sum user rate compared to conventional DRL method and GNN method, with lower training and inference cost. The framework supports adaptive deployment across large scale wireless communications scenarios, establishing a universal solution for intelligent radio environments.
	
\end{itemize}

The rest of this paper is organized as follows. Section II presents the system model and problem formulation. The optimization problem is formulated as a LLM problem in Section III. Section IV proposes a LLM backbone for RIMSA control. Section V provides simulation results and Section VI concludes the paper.

\section{System Model}

\begin{figure}[t]
	\centering
	\includegraphics[width=3 in]{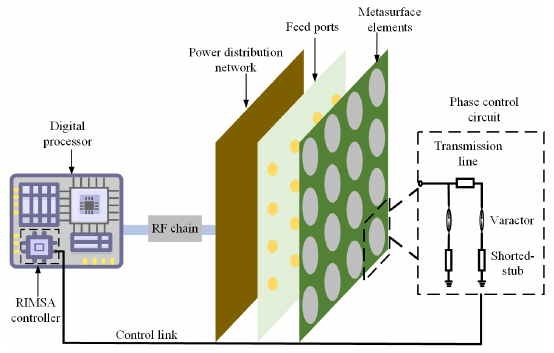}
	\caption{The architecture of RIMSA.}
	\label{RIMSA_structure}
\end{figure}

\subsection{RIMSA Architecture}
The RIMSA represents a highly adaptable metasurface antenna, capable of dynamic adjustment and reconfiguration in response to varying wireless environments. As depicted in Fig. \ref{RIMSA_structure}, the structure of a RIMSA incorporates phase-shifting circuits, metamaterial elements, a power distribution network, and feed ports. Each metamaterial element phase control circuit is tasked with modulating the phase of the electromagnetic (EM) signal, utilizing a varactor, transmission line, and short stub. By adjusting the control voltage applied to the phase control circuit, the capacitance of the varactors can be tuned, thereby altering the phase response of the corresponding metamaterial element. The phase-shifting mechanism depends on the phase control circuit, enabling continuous phase modulation of the EM wave and facilitating analog beamforming. The RF power is evenly distributed across all metamaterial elements via the power distribution network, which is realized using microstrip line power dividers. The feed ports include a primary feed port along with auxiliary ports for power division. These four core components are tightly integrated, ensuring a compact overall antenna design. All metamaterial elements are simultaneously excited by the incoming signal and subsequently directed to the RF chain and digital processor through a parallel feed network, mitigating frequency selectivity commonly encountered in antenna architecture of DMA and RHS.

Traditional hybrid analog-digital MIMO architectures typically depend on specialized analog combining hardware, such as phase shifter networks, which inherently consume significant power due to the active tuning demands of the phase shifters. In contrast, the radiation pattern of a RIMSA is generated through dynamically tunable metamaterial elements. These elements are embedded within a parallel feeding network, primarily composed of passive components like varactor diodes, leading to a marked reduction in power consumption. Additionally, the limited response speed of conventional phase shifters results in fixed radiation patterns during a single symbol period in traditional hybrid MIMO systems. However, the RIMSA achieves symbol-level beam reconfiguration, thanks to the sub-nanosecond response time of varactor diodes. This allows the metamaterial elements to be swiftly reconfigured by the control circuitry within a single symbol duration, enabling real-time adaptation to rapidly changing channel conditions.

\begin{figure}[t]
	\centering
	\includegraphics[width=3 in]{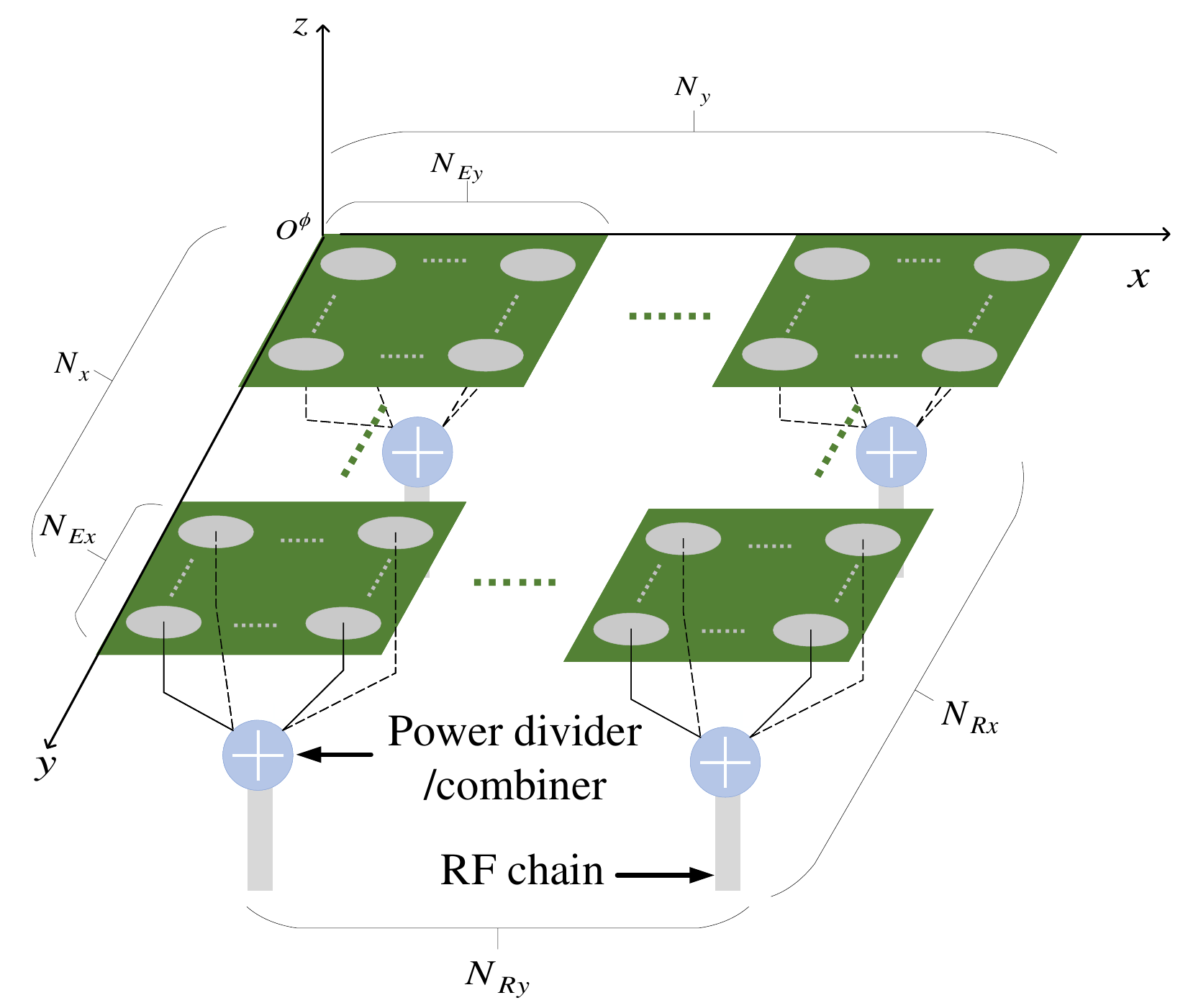}
	\caption{A planar RIMSA array with $N_{R_y} N_{R_x}$ antennas (RF chains) and a total of $N_{x} N_{y}$ metamaterial elements, where each RIMSA is a planar antenna as well with $N_{E_y} N_{E_x}$ elements.}
	\label{RIMSA_array}
\end{figure}

As illustrated in Fig. \ref{RIMSA_array}, a general RIMSA antennas consists of $N_{E}=N_{E_x} \times N_{E_y}$ densely spaced subwavelength metamaterial elements. The RIMSA array is composed of $N_{R}=N_{R_x} \times N_{R_y}$ RIMSAs connected to RF chains. The RIMSA array consists of $N_{t}=N_{E_x}N_{R_x} \times N_{E_y}N_{R_y} =N_{x} \times N_{y}$ in total elements.
Integrated electromagnetic wave control and reconfiguration is achieved through dynamic configuration of the metamaterial elements phase responses. For $n_r = 1,2 ...,N_R, n_e=1,2,...,N_E$, the effect of phase shifting of metamaterial elements is
$\mathbf{v}_{n_{r}, n_{e}}=e^{-j \alpha_{n_{r}, n_{e}}}$ which is the weighting factor of $n_{e}$th metamaterial element of $n_{r}$th RF chain. $\alpha_{n_{r}, n_{e}}$ denotes the phase shift. We define vector $\mathbf{v}_{n_{r}}$ as the phase shift vector of $n_{r}$th RF chain, which is
\begin{equation}
\mathbf{v}_{n_{r}}=1 / \sqrt{N_{E}}\left[v_{n_{r}, 1}, v_{n_{r}, 2}, \ldots, v_{n_{r}, N_{E}}\right]^{T}. \label{RFchain}
\end{equation}

\subsection{Signal Model}
We consider a MISO system that includes a base station (BS) and $K$ single-antenna users. The BS equipped with a RIMSA is made up of uniform planar array (UPA). The mobile user is equipped with an omnidirectional antenna as Fig. \ref{angle}.
The uplink channel characteristics between the base station and user $k$ are characterized by $\mathbf{h}_{k} \in \mathbb{R}^{N_{t} \times 1}$. 
In traditional methods, CSI acquisition precedes data transmission through uplink pilot signaling. Leveraging channel reciprocity, the downlink channels correspond to the transpose of uplink estimates. Specifically, user $k$ transmits a length-$L$ pilot sequence $x_{k}(\ell), \ell=1, \cdots, L $. The BS received signal in time slot $\ell$ is expressed as:
\begin{equation}
	\mathbf{y}(\ell)  =\sum_{k=1}^{K}\mathbf{V}^{H} \mathbf{h}_{k} x_{k}(\ell)+\mathbf{V}^{H}\mathbf{n}(\ell), 
\end{equation}
where $\mathbf{n}(\ell) \sim \mathcal{C N}\left(\mathbf{0}, \sigma_{1}^{2} \mathbf{I}\right)$ is the additive white Gaussian noise (AWGN). $\mathbf{V} \in \mathbb{C}^{N_{t} \times N_{R}}$ represents the beamforming matrix at the RIMSA array as:
\begin{equation}
	\mathbf{V}=\left[\begin{array}{cccc}
		\mathbf{v}_{1} & \mathbf{0} & \ldots & \mathbf{0} \\
		\mathbf{0} & \mathbf{v}_{2} & \ldots & \mathbf{0} \\
		\vdots & \vdots & \ddots & \vdots \\
		\mathbf{0} & \mathbf{0} & \mathbf{0} & \mathbf{v}_{N_{R}}
	\end{array}\right]
\end{equation}
where $\mathbf{v}_{n_r}$ is in the form of (\ref{RFchain}).
The unitary property $\mathbf{V}^{H}\mathbf{V} = \mathbf{I}$ ensures that the noise preserves its statistical properties, with its distribution and power remaining unchanged.
We assume channel reciprocity, so that the channel matrices in the downlink direction are the transpose of the uplink channels.

Let $s_{k} \in \mathbb{C}$ denote the modulated symbol transmitted from the BS to user $k$. The received signal at user $k$ is modeled as:
\begin{equation}
	r_{k}= \mathbf{h}_{k}^{H} \mathbf{V} \mathbf{w}_{k} s_{k} + \sum_{i \neq k} \mathbf{h}_{k}^{H} \mathbf{V} \mathbf{w}_{i} s_{i} + n_{k},
\end{equation}
where $n_{k} \sim \mathcal{C N}\left(0, \sigma_{0}^{2}\right)$ is AWGN, $\mathbf{w}_{i} \in \mathbb{C}^{N_{R} \times 1}$ is the digital precoding vector.

\subsection{Channel Model}
We assume that the RIMSA is deployed at a location where a line-of-sight channel exists between RIMSA and the users, so the Rician fading channel model $\mathbf{h}_{k}$ for the RIMSA-user $k$ link is expressed as:
\begin{equation}
	{\bf{h}}_{k} = \sqrt{L_{k}}\left(\sqrt{\frac{K}{K+1}} {\bf{h}}_{k,LoS}+\sqrt{\frac{1}{K+1}} {\bf{h}}_{k,NLoS}\right),
\end{equation}
where  $L_{k} \triangleq L_{1} d_{k}^{-\rho_{k}}$, $\frac{1}{L_{BE}}$ represents the large-scale path loss with $d_{k} \triangleq \| \mathbf{p}_{R}-\mathbf{p}_{k} \|$  denoting the RIMSA-user $k$ distance and $\rho_{k}$ being the path loss exponents. 
The Rician factor $K$ quantifies the power ratio between line-of-sight (LoS) and non-line-of-sight (NLoS) components. The deterministic LoS component $\mathbf{h}_{k, LoS}$ exhibits slow temporal variation due to positional stability, while the NLoS component $\mathbf{h}_{k, NLoS}$ denotes the fast fading NLoS component, and the component of which is independent and identically distributed (i.i.d.) circularly symmetric complex Gaussian random variables with zero-mean and unit variance.
The LoS component is position-dependent and slow-time-varying; The NLoS components are caused by the
multi-path effects and are thus fast-time-varying. 

\begin{figure}[t]
	\centering
	\includegraphics[width=3 in]{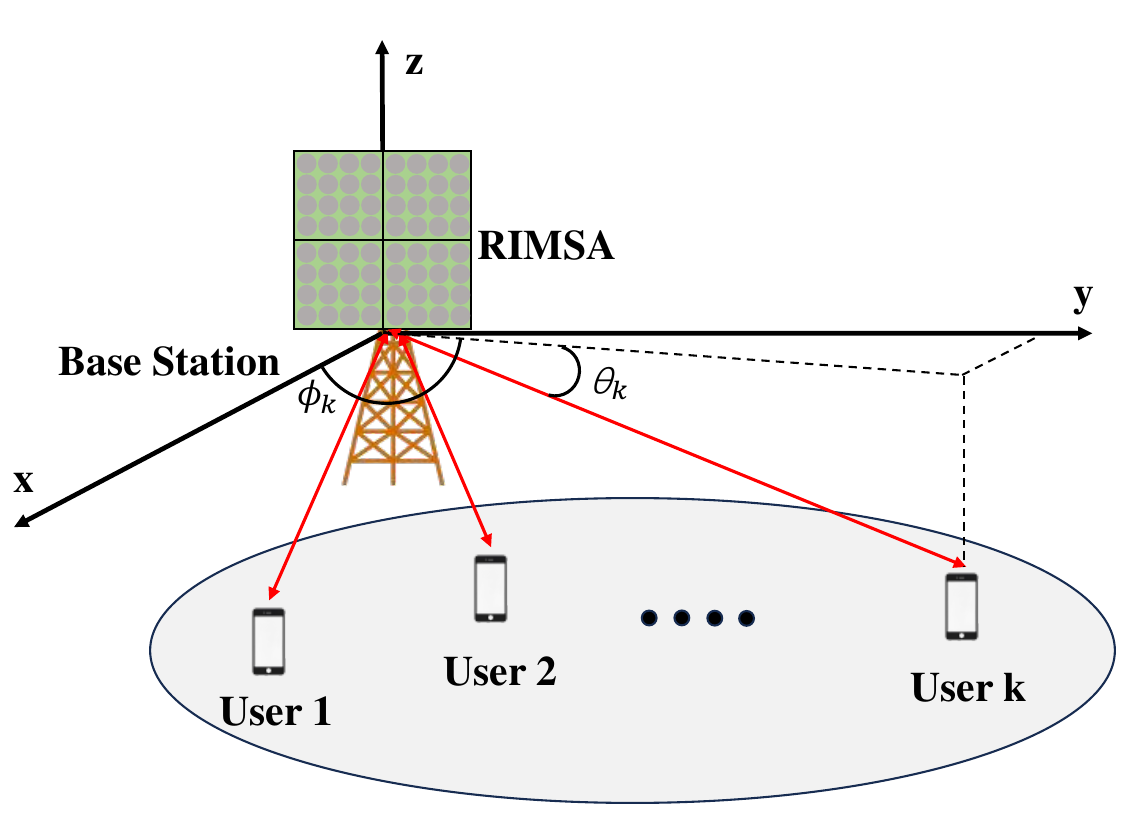}
	\caption{Simulation layout of RIMSA assisted multiuser system.}
	\label{angle}
\end{figure}

As illustrated in Fig. \ref{angle}, the LoS channel depends on user-RIMSA geometric relationships. Let $\phi_k$ and $\theta_k$ respectively denote the azimuth and elevation angles of arrival from user $k$ to RIMSA. The $n$-th element of the RIMSA steering vector ${\bf{a}}_{R}(\phi_k, \theta_k)$ is given by:
\begin{equation}
	\begin{array}{l}
		{\left[{\bf{a}}_{R}(\phi_k, \theta_k)\right]_{n}=} 
		\quad e^{j \frac{2 \pi d_{\mathrm{R}}}{\lambda_{c}}\left\{i_{1}(n) \sin \left(\phi_{k}\right) \cos \left(\theta_{k}\right)+i_{2}(n) \sin \left(\theta_{k}\right)\right\}},
	\end{array}
\end{equation}
where $d_{\mathrm{R}}$ represents the inter-element spacing, and $\lambda_{c}$ is the carrier wavelength, $i_{1}(n)=\bmod (n-1,\sqrt{N_t})$, and $i_{2}(n)=\lfloor(n-1) / \sqrt{N_t}\rfloor$ which is the index of $n$-th element. In the simulations, we assume  $d_{\mathrm{R}}=\frac{{\lambda_{c}}}{2}$ without loss of generality. This leads to the LoS channel expression:
\begin{equation}
	{\bf{h}}_{k,LoS}={\bf{a}}_{R}(\phi_k, \theta_k).
\end{equation}

Let $\left(x_{k}, y_{k}, z_{k}\right)$ denote the location of the user $k$ and $(\left.x_{\mathrm{R}}, y_{\mathrm{R}}, z_{\mathrm{R}}\right)$ denote the location of the RIMSA, we have
\begin{equation}
		\sin \left(\phi_{k}\right) \cos \left(\theta_{k}\right)  =\frac{y_{k}-y_{\mathrm{R}}}{d_{k}}, 
		\sin \left(\theta_{k}\right)  =\frac{z_{k}-z_{\mathrm{R}}}{d_{k}},
\end{equation}
where $d_{k}$ is the distance between RIMSA and the user $k$.

\section{Optimization Problem for LLM-RIMSA}
In this section, we introduce the optimization goal and loss function for RIMSA control of LLM.

\subsection{Optimization Problem}
Since the number of elements $N_{t}$ in a typical RIMSA is generally large (possibly in thousands), it is challenging to estimate the channels.
The main idea of this paper is that since the final goal is to optimize the rates $R_{k}$, instead of explicitly estimating all the channel coefficients as an intermediary step, we can exploit the pilot phase more efficiently by mapping the received pilots directly to the optimized transmission strategy for data rate maximization, in effect, bypassing channel estimation.

Under a block-fading channel model, channel parameters remain constant within coherence intervals but vary independently across blocks. 
The corresponding signal-to-noise ratio (SINR) metric becomes:
\begin{equation}
	\operatorname{SINR}_{k}=\frac{\left|\mathbf{w}_{k}^{H} \mathbf{V}^{H} \mathbf{h}_{k}\right|^{2}}{\sum_{i \neq k}^{K}\left|\mathbf{w}_{i}^{H} \mathbf{V}^{H} \mathbf{h}_{k}\right|^{2}+\sigma_{k}^{2}}.
\end{equation}

The achievable rate for user $k$ is computed as:
\begin{equation}
	R_{k}=\log \left(1+ \operatorname{SINR}_{k} \right), \label{datarateloss}
\end{equation}
where multiuser interference is treated as noise. The RIMSA optimizes digital precoding vectors $\mathbf{w}_{k}$ and RIMSA beamforming matrix $\mathbf{V}$ to maximize a network utility function $U\left(R_{1} \ldots R_{K}\right)$. Typical utility functions include system sum-rate $\sum_{k=1}^{K} R_{k}$.

The system-level performance is quantified through the sum-rate capacity:
\begin{equation}
	R_{sum}=\sum_{k=1}^{K} R_{k}=\sum_{k=1}^{K} \log _{2}\left(1+\operatorname{SINR}_{k}\right), \label{sumdata}
\end{equation}

The optimization for RIMSA-enhanced systems is to control RIMSA beamforming matrix $\mathbf{V} \in \mathbb{C}^{N_{t} \times N_{R}}$  and digital precoding
matrix $\mathbf{W} \in \mathbb{C}^{N_{R} \times K}$ to maximum the sum data rate ${R_{sum}}$. 
To this end, we propose to design the optimal beamforming vector $\mathbf{w}_{k}$ and RIMSA beamforming matrix $\mathbf{V}$ based directly on the received pilots  $\mathbf{Y}=[\mathbf{y}(1), \mathbf{y}(2), \cdots, \mathbf{y}(L)] \in \mathbb{C}^{{N}_R \times L}$. Specifically, given the matrix $\mathbf{Y}$, our goal is to solve the following optimization problem according to (\ref{datarateloss}):
\begin{equation}
	\begin{array}{ll}
		\underset{\mathbf{W},\mathbf{V} = g(\mathbf{Y})}{\operatorname{maximize}} & \mathbb{E}\left[R_{1}(\mathbf{W}, \mathbf{V}), \ldots, R_{K}(\mathbf{W}, \mathbf{V})\right] \\
		\text{\rm{ s}}{\rm{.t}} & \mathbf{W} = \left[{\mathbf{w}_{1}, \cdots, \mathbf{w}_{K}} \right],\\
		& {\| \mathbf{w}_{i} \|}^{2} \leq P_{max},             \\
		& \mathbf{V}^{H}\mathbf{V} = \mathbf{I}, \label{pilot}
	\end{array}
\end{equation}
where $\mathbf{W}$ is the digital precoding matrix of RIMSA. Solving problem (\ref{pilot}) 
faces critical challenges from two perspectives: 

(1) The large number of RIMSA elements (possibly in thousands) makes explicit channel estimation impractical due to prohibitive pilot overhead scaling. 

(2) Even with perfect CSI, the high-dimensional RIMSA configuration optimization involves a nonconvex combinatorial search space with complexity growing exponentially with $N_t$, rendering traditional convex relaxation methods computationally intractable. 

To address these dual challenges, we propose LLM-RIMSA to learn the mapping from pilot signals to optimal configurations through neural network parameter learning, bypassing explicit channel estimation while embedding physical constraints in the optimization process.


\subsection{Training Loss Configuration}
The proposed neural network is trained on pilot signals to control RIMSA with a hybrid loss function combining multiple optimization objectives. During the training phase, the hybrid loss function is formulated as follows:
\begin{equation}
	\begin{aligned}
		\mathcal{L}_{\text{total}} &= \mathcal{L}_{\text{mse}} + \lambda_{\text{rate}} \mathcal{L}_{\text{rate}} + \lambda_{\text{precoding}}\mathcal{L}_{\text{fro}}, 
	\end{aligned}
\end{equation}
where $\mathcal{L}_{\text{mse}}$ denotes the channel estimation MSE loss as:
\begin{equation}
	\mathcal{L}_{\text{mse}} = \|\hat{\mathbf{\bf{H}}}_{\mathrm{E}}-\mathbf{\bf{H}}_{\mathrm{truth}}\|_\mathrm{F}^2,
\end{equation}
where $\mathbf{\bf{H}}_{\mathrm{truth}}$ is the truth value of wireless channel. 
\footnote{In fact, an explicit channel estimation is only for training, the RIMSA configurations will be optimized by LLM model when receives the uplink pilot signals.}
$\mathcal{L}_{\text{rate}}$ reflects the sum data rate according to (\ref{sumdata}) as:
\begin{equation}
	\mathcal{L}_{\text{rate}} = -\sum_{k=1}^{K} R_{k}=-\sum_{k=1}^{K} \log _{2}\left(1+\operatorname{SINR}_{k}\right),
\end{equation}
where $\mathcal{L}_{\text{rate}}$ is determined by $\mathbf{V}$ and $\mathbf{W}$ according to (\ref{sumdata}), and $\mathcal{L}_{\text{rate}}$ is a negative value, because the LLM model will minimize the hybird loss. Being negative can help LLM model optimize RIMSA configuration to achieve maximum sum data rate. 
$\mathcal{L}_{\text{fro}}$ correspond to zero force (ZF) precoding matrix Frobenius norm loss as:
\begin{equation}
	\mathcal{L}_{\text{fro}} = \|\mathbf{M}-\mathbf{V}\mathbf{W}\|_\mathrm{F}^2 ,
\end{equation}
where $\mathbf{M} \in \mathbb{C}^{N_{R} \times K}$ is the ZF precoding matrix and $\mathcal{L}_{\text{fro}}$ can help LLM model faster optimize its parameters \cite{W2022Intelligent}.
$\lambda_{\text{rate}}$ is increasing with training epochs increases, so the optimization process will tend to maximum sum data rate.
In addition, the validation loss also adopts $\mathcal{L}_{\text{rate}}$ as loss function. It is worth noting that the multi-head attention and feed forward layers of the pre-trained GPT-2 are frozen, while other parameters of the network are trainable. Since the former includes the main parameters of the network, the number of trainable parameters is relatively small. The model with the smallest validation loss is saved for the testing phase.

\section{The LLM Model for RIMSA Control}
\begin{figure*}[t]
	\centering
	\includegraphics[width=6 in]{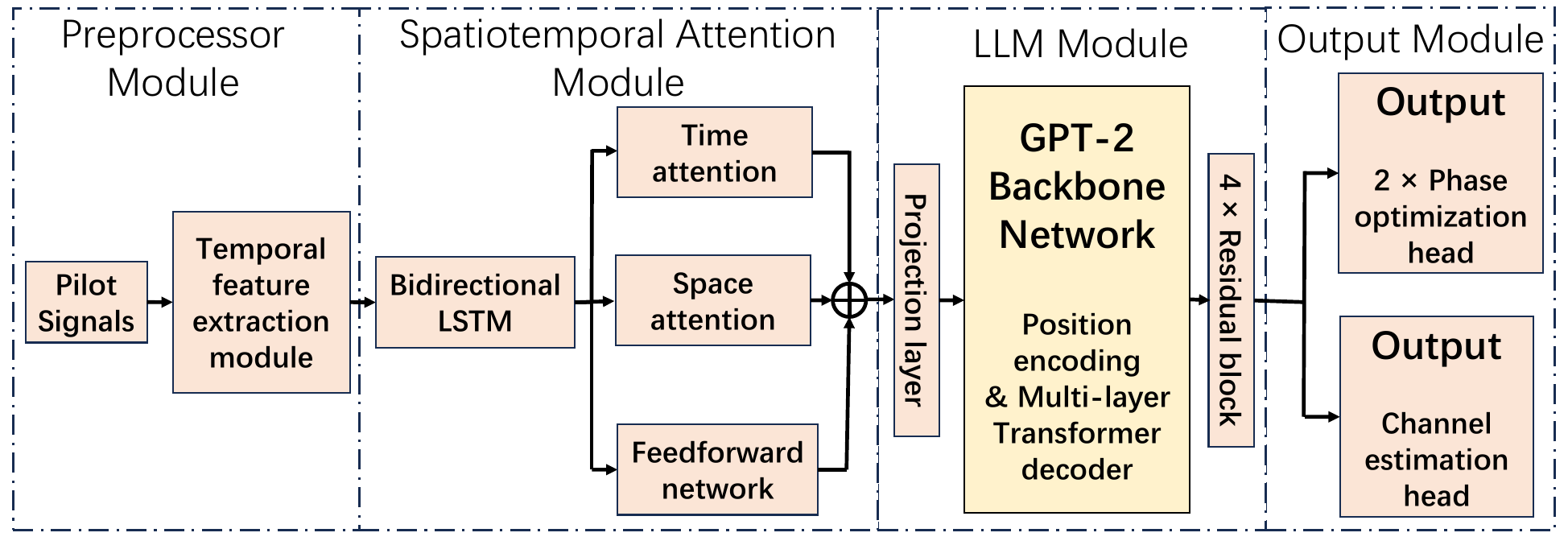}
	\caption{An illustration of the network architecture of LLM-RIMSA.}
	\label{gptmodel}
\end{figure*}

In this section, we introduce how to apply a pre-trained LLM model on RIMSA control. The proposed LLM-empowered RIMSA control framework integrates temporal feature extraction, spatio-temporal attention encoding, a GPT-based backbone, and joint multi-task outputs. The detailed implementation is structured in Fig. \ref{gptmodel}.

The proposed LLM-RIMSA framework establishes an end-to-end signal processing architecture through four synergistic components: Preprocessor module converts raw pilot signals into noise-robust temporal-spatial representations via multi-scale convolutional filtering and bidirectional sequence modeling; Spatio-temporal attention module decouples and refines joint signal dependencies using parallelized temporal correlation learning and spatial affinity modeling; LLM module leverages pretrained GPT-2 transformers with adaptive positional encoding and residual refinement blocks to generate high-dimensional semantic embeddings; Output module implements dual-task decoding with physical constraints enforcement, producing phase control matrices and channel estimation parameters through task-specific nonlinear projections. 
This unified architecture achieves multi-task optimization by progressively transforming pilot signals from raw measurements to control parameters through hierarchical feature abstraction and context-aware reasoning.

\emph{Remark 1}: Different from prompt-based related works, which focuses on integrating multimodal and enhancing domain knowledge through RAG and external tools, LLM-RIMSA emphasizes the design of specialized modules to adjust the model architecture to the input feature space of LLMs. Besides, LLM-RIMSA demonstrates distinct advantages in achieving high prediction accuracy and strong generalization performance.

\subsection{Preprocessor Module}

The temporal feature extraction module processes raw received signals through a composite convolutional architecture.
Given pilot signals with input dimensions $\mathbf{Y} \in \mathbb{C}^{N_{R} \times L}$  ($N_{R}$ is the number of RIMSA RF chains and $L$ is the length of pilot signals). So the input tensor can be expressed as 
\begin{equation}
	\mathbf{X}=\left[\begin{array}{l}
		\Re({\mathbf{\bf{Y}}}) \\
		\Im({\mathbf{\bf{Y}}})
	\end{array}\right] \in \mathbb{R}^{2 \times N_{R} \times L},
\end{equation}
where $\Re(\cdot)$ and $\Im(\cdot)$ denote the real and imaginary parts of the matrix, respectively. This results in a real-valued tensor $mathbf{X}$ with dimensions $2 \times N_{R} \times L$, where the first dimension separates the real and imaginary components. Reshape $\mathbf{X}$ from dimensions $2 \times N_{R} \times L$ to $2N_{R} \times L$, i.e., $\mathbf{X} \in \mathbb{R}^{2N_{R} \times L}$.

The preprocessor module performs temporal feature extraction and bidirectional sequence modeling to generate noise-robust representations for downstream modules.

A wide-kernel convolution $\mathrm{Conv1D}$ where $\mathrm{C_{out}}=2N_{R}$ and $\mathrm{kernel=3}$ expands the feature dimension while capturing local temporal correlations:
\begin{equation}
\mathbf{X} \in \mathbb{R}^{2N_{R} \times L} \xrightarrow{\text { Conv1D }} \mathbf{X}_{Conv} \in \mathbb{R}^{2N_{R} \times 2N_{R}}   .
\end{equation}
Then, the Gaussian error linear unit (GELU) activation and batch normalization (BN) enhance nonlinear expressivity and stabilize training. Max pooling reduces temporal redundancy by half, prioritizing dominant features:
\begin{equation}
\mathbf{X}_{Conv} \in \mathbb{R}^{2N_{R} \times 2N_{R}} \xrightarrow{\text { MaxPool1D(2) }} \mathbf{X}_{MaxPool} \in \mathbb{R}^{N_{R} \times 2N_{R}}    .
\end{equation}

A bidirectional LSTM (BiLSTM) captures long-range temporal dependencies due to its ability of contextual awareness and sequence alignment.
After temporal feature extraction, A bidirectional LSTM with hidden size $N_{R}$ processes the compressed sequence $\mathbf{X}_{MaxPool}$, capturing long-range dependencies in both forward and backward directions:
\begin{equation}
\mathbf{X}_{MaxPool} \in \mathbb{R}^{N_{R} \times 2N_{R}} \xrightarrow{\text { BiLSTM }} \mathbf{X}_{BiLSTM} \in \mathbb{R}^{N_{R} \times 2N_{R}}    .
\end{equation}

Each direction outputs $\mathbf{X}_{BiLSTM}$, concatenation preserves the full temporal resolution while doubling feature richness. Forward/backward hidden states concatenate to form $2N_{R}$-dim embeddings, encoding both historical and future context. Input features are permuted for temporal processing. Residual connections between convolutional and LSTM outputs mitigate vanishing gradients.

The key design rationale behind this architecture emphasizes multi-scale feature fusion where the Conv1D-BN-Pooling chain hierarchically extracts local-to-global patterns and the BiLSTM bridges distant timesteps to mitigate signal distortion in noisy channels. Ensuring dimension consistency, the output size aligns perfectly with the spatio-temporal attention module input requirements, facilitating seamless feature propagation. This preprocessing flow transforms raw pilot signals into temporally coherent and spatially enriched embeddings. These embeddings form the foundation for subsequent attention-based refinement and LLM-driven semantic modeling, enhancing the overall robustness and effectiveness of signal processing in complex communication environments.

\subsection{Spatio-temporal Attention Module}
As the important component of the architecture, the spatio-temporal attention module processes the bidirectional LSTM outputs $\mathbf{X}_{BiLSTM}$ and learns joint spatio-temporal dependencies through a decoupled dual-attention mechanism.
A novel multi-head attention architecture addresses spatial-temporal correlations. Temporal attention computes dependencies across time steps using scaled dot-product attention. Spatial attention operates on transposed features to model inter-antenna relationships. Attention outputs are added to original features via residual connections, preserved through layer normalization. In the meanwhile, dimension permutation avoids explicit tensor reshaping, reducing memory overhead.

Time attention (8-head) captures temporal dynamics across $N_{R}$ meta-elements using multi-head scaled dot-product attention:
\begin{equation}
\operatorname{Attention}\mathbf{(Q, K, V)}=\operatorname{Softmax}\left(\frac{\mathbf{Q} \mathbf{K}^{\top}}{\sqrt{d_{T}}}\right) \mathbf{V},
\end{equation}
where $\mathbf{Q / K / V} \in \mathbb{R}^{0.5N_{R} \times 0.25N_{R}}$ are split from $\mathbf{X}_{BiLSTM}$. The 8-head outputs are concatenated and projected to maintain dimension $\mathbb{R}^{N_{R} \times 2N_{R}}$.

Space attention (8-head) models inter-feature relationships among $2N_{R}$ channels via parallel spatial attention heads. And each head computes:
\begin{equation}
\text { Attention }_{\text {space }}=\sigma\left(\mathbf{W}_{s} \cdot \mathbf{X}_{BiLSTM}^{\top}\right) \cdot \mathbf{X}_{BiLSTM},
\end{equation}
where $\mathbf{W}_{s}$ is a learnable spatial affinity matrix. The outputs are fused to preserve $\mathbb{R}^{0.5N_{R} \times 2N_{R}}$.

Feedforward network (FFN) applies two linear layers with GELU activation and four hidden dimension expansion as:
\begin{equation}
\mathbf{X}_{BiLSTM} \in \mathbb{R}^{N_{R} \times 2N_{R}} \xrightarrow{\mathrm{FFN}} \mathbf{X}_{FFN} \in \mathbb{R}^{N_{R} \times 2N_{R}}   .
\end{equation}

Feature fusion combines temporal and spatial attention outputs through residual addition as:
\begin{equation}
\mathbf{X}_{\text {fused }} = \mathbf{X}_{\text {time }} + \mathbf{X}_{\text {space }} + \mathbf{X}_{\text {space }}.
\end{equation}

The design of this module adopts a decoupled spatio-temporal learning approach, where the separation of time and space attention reduces computational complexity while enabling targeted feature refinement. This is complemented by a dynamic weight allocation mechanism, utilizing a dual-head structure to adaptively prioritize critical timesteps, such as sudden channel fading, and spatial features, like antenna-specific distortions. Ensuring dimension preservation, the output maintains consistent dimensions of $\mathbf{X}_{\text {fused }} \in \mathbb{R}^{2N_{R} \times L}$, facilitating seamless integration with the projection layer and GPT-2 backbone. Illustrated in Fig. \ref{gptmodel}, this module generates context-aware spatio-temporal embeddings that encapsulate both global temporal trends and local spatial correlations. These embeddings provide essential inductive biases for phase prediction and channel estimation tasks, enhancing the accuracy and efficiency of communication systems.

\subsection{LLM Module}
\begin{figure}[t]
	\centering
	\includegraphics[width=2 in]{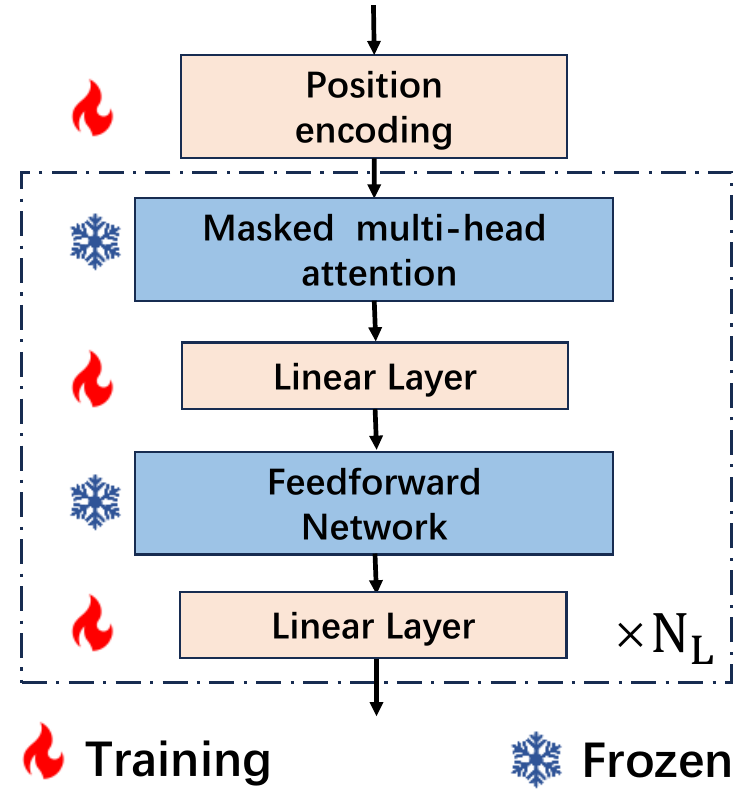}
	\caption{An illustration of the network architecture of GPT-2 backbone network.}
	\label{llmmodel}
\end{figure}

The LLM Module serves as the semantic reasoning core of the architecture, transforming spatio-temporal features into high-dimensional contextual representations. There are different kinds of GPT-2 models and we apply normal GPT-2 which consists of millions parameters into this problem. This module comprises three key components: learnable position encoding, $N_L$-layer transformer decoder and 4-stage residual block group ($N_L$ is set as 6 there). The details of GPT-2 backbone network is shown in Fig. \ref{llmmodel}.

\begin{table}[t]
	\centering  
	\captionsetup{justification=raggedright, singlelinecheck=false}
	\caption{
		GPT-2 Model Dimension Specifications
	}
	\label{gptdimension}
	\begin{tabular}{@{}lccc@{}}
		\toprule
		\textbf{Model Name} & \textbf{Type Parameter} & \textbf{Feature Dimension}  \\ \midrule
		GPT-2 Base        & gpt2              & 768 ($N_P = 768$)                                 \\
		GPT-2 Medium      & gpt2 medium       & 1024 ($N_P = 1024$)                               \\
		GPT-2 Large       & gpt2-large        & 1280 ($N_P = 1280$)                               \\
		GPT-2 XL          & gpt2-xl           & 1600 ($N_P = 1600$)
		\\ \bottomrule
	\end{tabular}
\end{table}

Before learnable position encoding, $\mathbf{X}_{\text {proj}} \in \mathbb{R}^{N_P}$ is generated by $\mathbf{X}_{\text {fused }}$ through projection layer. The learnable position encoding injects temporal order information into the projected features as:
\begin{equation}
	\mathbf{X}_{\mathrm{pos}}=\mathbf{X}_{\mathrm{proj}}+\mathbf{W}{p} \cdot \mathbf{P}_{\mathrm{enc}},
\end{equation}
where $\mathbf{P}_{\text {enc }} \in \mathbb{R}^{N \times N_P}$ is a trainable positional embedding matrix according to Table \ref{gptdimension}, and $N$ denotes the maximum sequence length.

Then the 6-layer transformer decoder process $\mathbf{X}_{\mathrm{pos}}$, each layer contains 12-head masked attention and 4 expanded feedforward networks. The 12-head causal attention processes sequences with auto-regressive masking to prevent future information leakage as:
\begin{equation}
\operatorname{Attention}\mathbf{(Q, K, V)}=\text { MaskedSoftmax }\left(\frac{\mathbf{Q} \mathbf{K}^{\top}}{\sqrt{d_{L}}}\right) \mathbf{V},
\end{equation}
where $d_{L}=N_P / 12$ according to Table \ref{gptdimension}, multi-head outputs are concatenated and linearly projected.
4 expanded feedforward networks apply dimension expansion and compression with GELU activation:
\begin{equation}
\operatorname{FFN}(\mathbf{x})=\mathbf{W}_{2} \cdot \operatorname{GELU}\left(\mathbf{W}_{1} \cdot \mathbf{x}+\mathbf{b}_{1}\right)+\mathbf{b}_{2},
\end{equation}
where $\mathbf{W}_{1} \in \mathbb{R}^{N_P \times 2N_P}$ and $\mathbf{W}_{2} \in \mathbb{R}^{2N_P \times N_P}$ are the trainable parameters.

The 4 stage residual block group enhances local feature interactions through dual-branch convolution and squeeze-excitation (SE) channel attention. The dual-branch convolution parallel depth-wise convolutions (DWConv) capture multi-scale spatial patterns as
\begin{equation}
	\mathbf{X}_{\mathrm{conv}}=\mathrm{DWConv}_{3 \times 3}\left(\mathbf{X}_{\mathrm{trm}}\right) \oplus \mathrm{DWConv}_{3 \times 3}^{\prime}\left(\mathbf{X}_{\mathrm{trm}}\right).
\end{equation}

SE channel attention recalibrates feature channels via squeeze-excitation as:
\begin{equation}
\mathbf{X}_{\mathrm{se}}=\mathbf{X}_{\mathrm{conv}} \cdot \sigma\left(\mathbf{W}_{c} \cdot \operatorname{GAP}\left(\mathbf{X}_{\text {conv }}\right)\right).
\end{equation}

The key design rationale of this module focuses on hierarchical abstraction achieved through a 6-layer Transformer that progressively builds global context via stacked self-attention mechanisms, while residual blocks are employed to refine local signal characteristics. Ensuring dimension compatibility, the projection layer transforms spatio-temporal features from $\mathbb{R}^{2N_{R} \times L}$ to $\mathbb{R}^{N_P}$, effectively bridging these features with the GPT-2 input space. Leveraging pretraining compatibility, the module inherits GPT-2 pretrained weights for its transformer layers, facilitating effective transfer learning in wireless communication scenarios. Consequently, this module generates semantic-enriched embeddings $\mathbb{R}^{N_P}$, which encapsulate long-term temporal dependencies from pilot sequences, spatial correlations across antenna arrays, and channel-specific physical constraints. These enriched embeddings are then directed towards dual-output heads for phase optimization and channel estimation, thereby completing the end-to-end signal processing chain.

The proposed architecture employs GPT-2 as its foundational backbone, comprising 6 stacked transformer decoders with dimension-compatible modifications. The positional embedding layer is reinitialized as trainable parameters to accommodate wireless sequence lengths; Each decoder layer preserves frozen 12-head attention and 4 expanded FFN from pretrained weights, while enabling fine-tuning of layer normalization parameters and residual connections; The 4-stage residual group with dual-branch DWConv and SE attention is appended as trainable extensions. This configuration maintains $N_P$-dimensional hidden states throughout the processing chain, ensuring compatibility with alternative LLM backbones (e.g., Llama) through dimension-preserving adapter layers when considering computation-performance trade-offs.

\subsection{Output Module}
The Output Module operates as the task-specific decoder, generating physically constrained predictions for phase correction and channel estimation. As illustrated in Fig. \ref{gptmodel}, this dual-head architecture processes the LLM backbone outputs $\mathbf{X}_{\text {LLM}} \in \mathbb{R}^{N_P}$ through parallel task branches.

Phase optimization head consists of feature expansion, nonlinear processing and phase constraint. Feature expansion expands latent space dimensions via fully-connected layer:
\begin{equation}
\begin{array}{c}
	\mathbf{X}_{\text {phase }}^{(1)}=\mathbf{W}_{p} \cdot \mathbf{X}_{\text {LLM}}+\mathbf{b}_{p}, \quad\left(\mathbf{W}_{p} \in \mathbb{R}^{N_P \times 1024}\right). \\
\end{array}
\end{equation}

Nonlinear processing applies GELU activation and layer-norm for stable gradient flow:
\begin{equation}
\mathbf{X}_{\text {phase }}^{(2)}=\operatorname{LayerNorm}\left(\operatorname{GELU}\left(\mathbf{X}_{\text {phase }}^{(1)}\right)\right) \in \mathbb{R}^{1024}.
\end{equation}

Phase constraint enforces physical limits $(-\pi , \pi)$ through Hardtanh:
\begin{equation}
\mathbf{V}=\operatorname{Hardtanh}\left(\mathbf{X}_{\text {phase }}^{(2)}\right) \cdot \pi \in \mathbb{R}^{N_{\mathrm{t}}}.
\end{equation}
Similar to RIMSA beamforming matrix $\mathbf{V}$, the digital precoding matrix $\mathbf{W}$ is generated from second phase optimization head with power constraint.

Channel estimation head consists of high-dimensional mapping, LeakyReLU activation and dimension reshaping. High-dimensional mapping projects features to 2048D space for detailed reconstruction as:
\begin{equation}
\begin{array}{c}
	\mathbf{X}_{\text {channel }}^{(1)}=\mathbf{W}_{c} \cdot \mathbf{X}_{\text {LLM}}+\mathbf{b}_{c}, \quad\left(\mathbf{W}_{c} \in \mathbb{R}^{N_P \times 2048}\right). \\
\end{array}
\end{equation}

LeakyReLU activation preserves negative gradient information as:
\begin{equation}
\mathbf{X}_{\text {channel }}^{(2)}=\operatorname{LeakyReLU}\left(\mathbf{X}_{\text {channel }}^{(1)}\right) \in \mathbb{R}^{2048}.
\end{equation}

Dimension reshaping reconstructs complex channel matrix as:
\begin{equation}
\hat{\bf{H}}_E=\operatorname{Reshape}\left(\mathbf{X}_{\text {channel }}^{(2)}\right) \in \mathbb{R}^{2 \times N_{\mathrm{t}} \times K}.
\end{equation}
where the first dimension separates real/imaginary components.

The design of this module incorporates dual-head specialization to efficiently handle complex signal processing tasks. Both heads share the LLM-derived semantic features $\mathbf{X}_{\text {LLM}}$, promoting joint learning of correlated tasks, while engaging in task-specific processing tailored to their objectives: the phase head emphasizes signal periodicity through GELU+LayerNorm, whereas the channel head focuses on matrix completeness via high-dimension expansion. Physical constraints embedding is achieved by applying Hardtanh in the phase head to enforce physical limits $(-\pi , \pi)$ into $\mathbf{V}$, aligning with RF circuit limitations, and $\mathbb{R}^{2 \times N_{\mathrm{t}} \times K}$ reshaping is used to reconstruct complex channel matrices. An activation strategy utilizing LeakyReLU (with $\alpha = 0.01$) prevents gradient death during deep fading scenarios in channel estimation, and LayerNorm ensures stable training despite varying pilot signal magnitudes. This module finalizes the end-to-end processing chain, converting neural network outputs into engineering-actionable parameters for channel estimates, phase shifting matrix and beamforming matrix for RIMSA systems.



\section{Numerical Results}
In this section, we evaluate the performance of the proposed LLM method for the RIMSA control in comparison to the channel estimation based approach and existing traditional deep learning methods.

\subsection{Simulation Parameters}

We consider a RIMSA assisted multiuser MISO communication system as illustrated in Fig. \ref{angle}, consisting of a BS equipped with a RIMSA which has 1024 elements and 128 RF chains. As shown in the Fig. \ref{angle}, the (x; y; z)-coordinates of the BS locations in meters is (0, 0, 20). There are 3 users uniformly distributed in a rectangular area
$[20, 30] \times [20, 30]$ in the (x, y)-plane with z = 1.5 as shown in Fig. \ref{angle}. We assume that the RIMSA is equipped with a uniform planar array placed on the (y, z)-plane in a $32 \times 32$ configuration, the RF chains are placed similar to RIMSA on the (y, z)-plane in a $16 \times 8$ configuration.
$K$ is the ratio between the power in the LoS path and the power in the NLoS paths.
The LoS component is position-dependent and is thus slow-time-varying; The NLoS components are caused by the
multi-path effects and are thus fast-time-varying. The large-scale path loss of $L_{k}$ is $10^{-4}$. The path loss exponents are set as $\rho_{k}=2.2$.
The uplink pilot transmit power and the downlink data transmit power are set to be 15dBm and 20dBm, and the uplink noise power is -100dBm and the downlink noise power is -80dBm unless otherwise stated.
The training dataset and validation dataset respectively contain 10,000 and 2000 samples. The testing dataset contains 5000 samples.

This work introduces a hybrid optimization network architecture based on GPT-2, designed to capture spatio-temporal channel evolution patterns and optimize phase, channel estimation, and precoding matrices through a multi-task learning framework. Specifically configured with an input preprocessing module for complex feature fusion and temporal feature extraction using dilated causal convolutions, it also includes a spatio-temporal joint encoding module featuring bidirectional LSTM for capturing long-term dependencies and a hybrid attention mechanism for processing sequence and feature dimensions. The backbone network adopts a customized GPT-2 configuration with parameters fine-tuned only for position encoding and layer normalization, while employing a multi-task output head strategy that encompasses phase prediction, channel estimation, and precoding prediction, each tailored for specific tasks like phase range constraint and spatial continuity smoothing. Additionally, a residual optimization module enhances the architecture with cascaded residual blocks and channel attention. This model, comprising tens of million parameters, achieves end-to-end latency of less than 20ms on an NVIDIA RTX4090 GPU, demonstrating efficient inference performance. Through its comprehensive design, this architecture effectively integrates cross-modal feature alignment and spatio-temporal dynamics for communication systems.

The model optimization employs a multi-stage training strategy with key hyper parameters as follows: The optimizer is AdamW ($\beta_1=0.9$, $\beta_2=0.999$), utilizing gradient accumulation over 4 steps to achieve an effective batch size of 4096. The learning rate starts at $3\times10^{-4}$ and is dynamically adjusted using a triangular OneCycle scheduling method, which varies the rate between $3\times10^{-4}$ and $1\times10^{-5}$ to balance convergence speed and stability. Regularization techniques include a weight decay of $1\times10^{-6}$ for ${l}_{2}$-norm penalty and gradient clipping with a threshold of 1.0 to prevent exploding gradients. The training protocol spans 500 epochs, with early stopping triggered if the validation SINR plateaus for 20 consecutive epochs. Mixed-precision training (FP16/FP32) is implemented via PyTorch AMP (Automatic Mixed Precision) to enhance memory efficiency. This configuration ensures stable optimization of large-scale parameters while preserving the physical constraints intrinsic to RIMSA-assisted communication systems.

To validate the superiority of the proposed method, several model-based and deep learning-based channel prediction methods are implemented as baselines as follows:
\begin{itemize}
	\item [$\bullet$]DRL \cite{Z2022Learning}: A deep deterministic policy gradient (DDPG) framework, featuring a dual-network architecture (actor-critic) for continuous phase-shift optimization. The actor network uses three fully connected layers to map pilots to RIMSA configurations.
	\item [$\bullet$]GNN \cite{T2021Learning}: A graph attention network (GAT) models RIMSA elements as nodes in a spatial graph. Each node aggregates features from 4 nearest neighbors, using three GAT layers with hidden dimension as 64 to capture interference correlations. 
	\item [$\bullet$]No optimization: RIMSA configurations are set randomly serve as the baseline.
\end{itemize}

To ensure fairness, the above deep learning-based methods process antenna dimensions in parallel and adopt sum data rate as the loss function for training.

\begin{figure}[t]
	\centering
	\includegraphics[width=3 in]{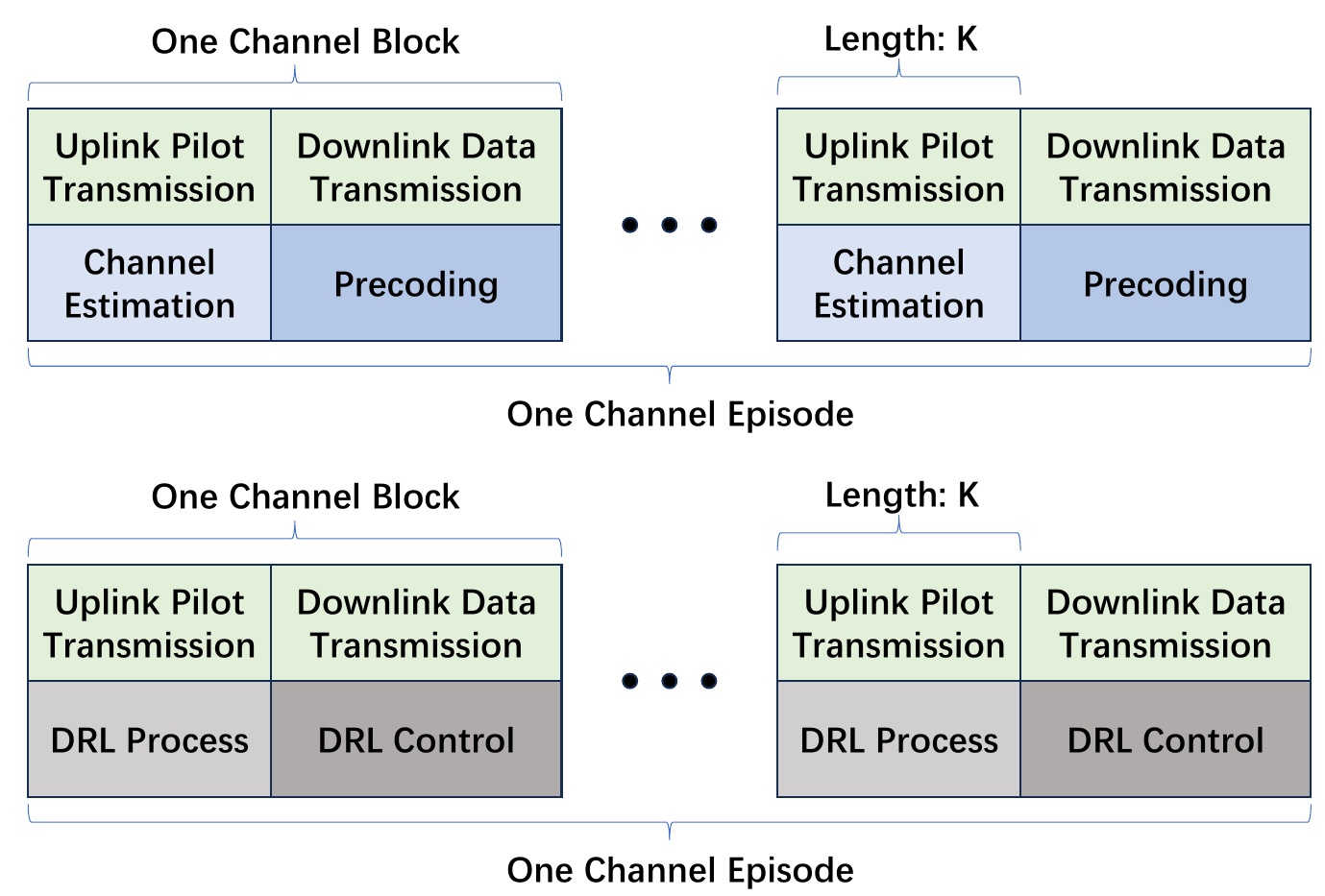}
	\caption{The frame structure of the typical TDD multi-user MIMO and the corresponding signal processing procedures for DRL method.}
	\label{tdd}
\end{figure}

\subsection{Training and Inference Cost}

In this subsection, we show the training and inference cost of LLM, GNN, and DRL in controlling the RIMSA.

The proposed GNN-based framework for RIMSA optimization adopts a multi-head GAT architecture to model spatial correlations among RIMSA elements. The framework contains three key components: a geometric graph constructor that generates adjacency matrices based on RIMSA element coordinates with $32 \times 32$ planar array and $16 \times 8$ RF chains, connecting each element to its horizontal/vertical neighbors; hybrid node features combining real/imaginary channel components across 3 users into 6-dimensional feature vectors; a three-layer GAT with 3-head attention and hidden dimension as 64 for hierarchical spatial feature extraction. The first GAT layer applies LeakyReLU activated multi-head attention to project features into 768-D space, followed by ReLU-activated and linear GAT layers that progressively aggregate neighbor information. The final phase prediction module uses batch normalization and tanh-activated dense layers to map node embeddings to $[-\pi, \pi]$ phase shifts, enabling end-to-end optimization of RIS configurations. 

The framework introduces two novel design principles: First, a system-level loss function directly maximizes the sum data rate through differentiable ZF precoding simulation. Given predicted phases, the loss computes equivalent channels, derives ZF precoding matrices via regularized matrix inversion, and evaluates SINR-based spectral efficiency through automatic differentiation. Second, a dynamic graph attention mechanism adaptively reweights neighbor connections based on channel state differences - elements with highly correlated channels receive stronger attention coefficients. Training employs Adam optimization with batch size as 32, using sum data rate as the primary metric. The adjacency matrix sparsity (4 connections/element) ensures real-time operation with about 50 ms inference latency per RIS configuration.

The LLM and GNN methods can directly get the optimal RIMSA configurations, while DRL method needs explorations to learn how to control RIMSA. So the training dataset for DRL is a little different, but the testing dataset is same as LLM and GNN methods. In DRL simulation based on Fig. \ref{tdd}, we make two assumptions:
Channel block contains uplink pilot transmission and downlink data transmission phases with constant channel matrices. Channel episode comprises 20 blocks where the LoS component (fixed via random user placement) remains static, while NLoS components vary as i.i.d. processes across blocks. This setup balances environmental stability with dynamic NLoS variations.

The proposed DRL-based optimization framework for RIMSA systems adopts a DDPG architecture to address the continuous phase-shift optimization problem. The framework consists of three core components: a channel-aware environment that models RIMSA dynamics, an actor-critic network for joint policy learning and value estimation, and an experience replay mechanism to stabilize training. The state space is constructed from received pilot signals and estimated channel matrices obtained through LMMSE estimation using orthogonal DFT pilots, concatenating real and imaginary components into a 768-dimensional vector. The action space defines phase shifts for all 1024 elements and 128 RF chains, bounded within $[-\pi, -\pi]$ through a tanh-activated actor network. The reward function maximizes the sum data rate by the agent phase-shift decisions on RIMSA.

To ensure robust learning, a hybrid exploration strategy combines warm-up random sampling (1,000 steps) with temporally correlated Ornstein-Uhlenbeck noise injection, balancing exploration-exploitation trade-offs in high-dimensional action spaces. And the network architecture employs dual hidden layers with 1024 neurons each and ReLU activation in both actor and critic networks, while target networks are softly updated to mitigate training divergence. Training leverages mini-batch SGD with batch size as 64 and Adam optimizers where actor/critic learning rates are set as $10^{-4}/10^{-3}$, minimizing the Bellman error via MSE loss on predicted Q-values. Experimental validation confirms convergence within 5,000 episodes, achieving real-time inference latency 120 ms on GPU hardware.

\begin{table}[t]
	\centering  
	\captionsetup{justification=raggedright, singlelinecheck=false}
	\caption{
		The deployment cost of different methods on RIMSA control
	}
	\label{deployment_cost}
	\begin{tabular}{@{}lccc@{}}
		\toprule
		\textbf{Method} & \textbf{Parameters} & \textbf{Training time} & \textbf{Interference time} \\ \midrule
		LLM      & 1.73/84.78 M       & 2.6h                            & 20ms                            \\
		GNN      & 2.12/2.12 M        & 4.8h                            & 50ms                            \\
		DRL      & 2.36/2.36 M        & 7.3h                            & 120ms                            
		\\ \bottomrule
	\end{tabular}
\end{table}

\begin{table}[t]
	\centering 
	\captionsetup{justification=raggedright, singlelinecheck=false}
	\caption{
		The details of LLM with different numbers of GPT-2 layers
	}
	\label{llm_cost}
	\begin{tabular}{@{}lccc@{}}
		\toprule
		\textbf{Layers} & \textbf{Sum datarate} & \textbf{Parameters} & \textbf{Interference time} \\ \midrule
		LLM (2)      & 7.01 bps/Hz        & 1.72/55.37 M                & 13ms                            \\
		LLM (4)      & 9.62 bps/Hz        & 1.73/70.26 M                & 17ms                            \\
		LLM (6)      & 9.79 bps/Hz        & 1.73/84.78 M                & 20ms                            \\
		LLM (8)      & 9.73 bps/Hz        & 1.74/99.12 M                & 28ms                            \\ \bottomrule
	\end{tabular}
\end{table}

We compare the model training and inference cost of the proposed method with other baselines to assess the difficulty of deploying the model in RIMSA control, as shown in Tab. \ref{deployment_cost}. All experiments
are conducted on the same machine with Intel Core i9-14900K CPU, NVIDIA GeForce RTX4090 GPU, and
128 GB of RAM. Although LLM has the largest total parameters as 84.78M, the actual trainable parameters 1.73M are similar to GNN and DRL models since most parameters in the LLM are frozen. And LLM method also has the shortest training time as 2.6 hours. It is worth noting that LLM inference time is shorter than GNN and DRL thanks to inference acceleration specific to the GPT model. Therefore, LLM has the potential to serve real-time RIMSA control.

In addition, we comprehensively evaluated the impact of selecting different numbers of GPT-2 layers on RIMSA control performance, parameters cost, and inference time, as shown in Tab. \ref{llm_cost} where LLM ($n$) represents the proposed method with $n$ GPT-2 layers. The network parameters and inference time are both increased with the number of GPT-2 layers. It is worth noting that the proposed model with 6 GPT-2 layers performs the best within the testing range, indicating that having more layers does not necessarily favor prediction. In practical deployment, the selection of the type and size of the LLM backbone needs to consider both the requirements for system performance and the constraints of device storage and computational resources.

\subsection{Performance Evaluation}
In this subsection, we show numerical results on the performance of RIMSA assisted wireless communications controlled by LLM model. In this experiment, the number of training samples is set as 10,000, the number of validate samples is set as 1,000 and the number of testing samples is set as 2,000.

\begin{figure}[t]
	\centering
	\includegraphics[width=3 in]{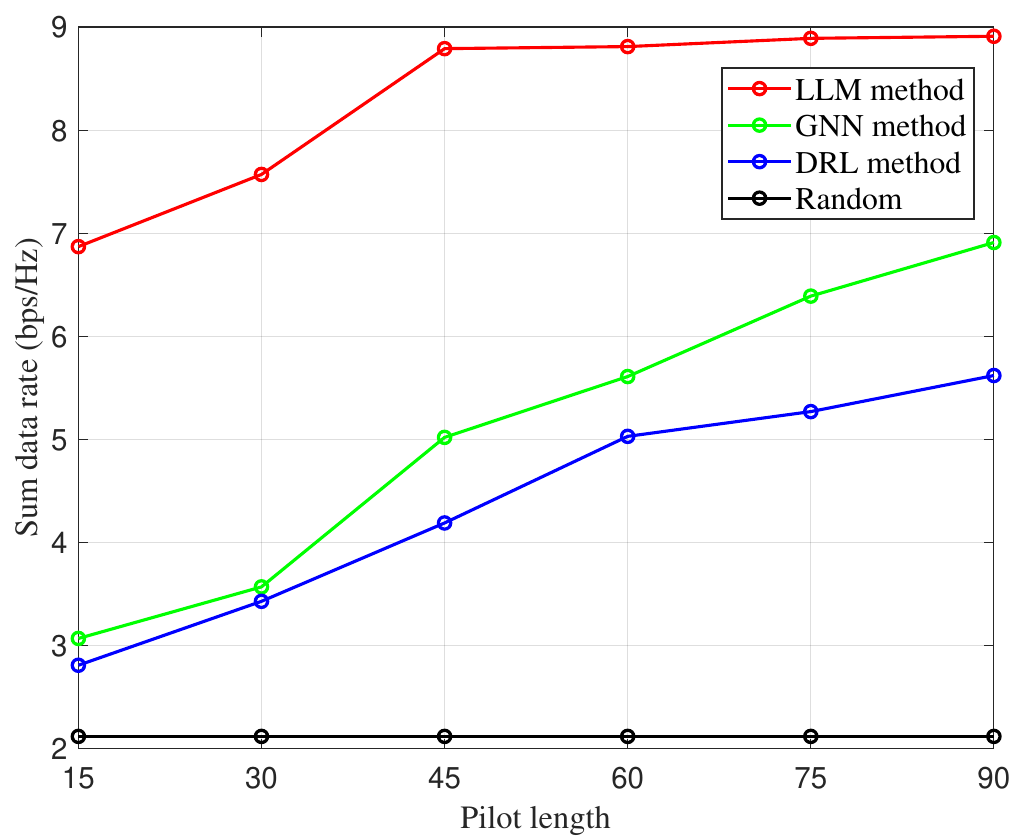}
	\caption{Performance of the proposed methods for RIMSA system with different pilot length.}
	\label{sumrate_pilot}
\end{figure}
The Fig. \ref{sumrate_pilot} evaluates the sum data rate (bps/Hz) of four methods LLM, GNN, DRL, and random under varying pilot lengths in a large-scale RIMSA environment. Notably, prior studies have not rigorously tested DRL or GNN in such complex multi-user, multi-RIMSA-element scenarios, which introduce unique challenges due to high-dimensional state and action spaces. Among these methods, GNN demonstrates a significant advantage through its graph-structured modeling, capturing spatial correlations between distributed RIMSA elements and users, thereby efficiently handling interference and channel state variations even with numerous RIMSA elements. At longer pilot lengths, GNN achieves near-optimal performance, as extended pilots provide richer node interaction data for graph-based optimization. In contrast, DRL struggles due to the exponentially growing state/action spaces (e.g., $\mathcal{O}(K^{N_t})$ for $K$ users and $N_t$ RIMSA elements), which impedes policy convergence, resulting in subpar performance. This curse of dimensionality renders DRL trial-and-error exploration computationally intractable, failing to adapt to dynamic multi-element interactions. The critical role of pilot length is evident: for GNN, longer pilots enhance its ability to resolve multi-user interference through high-resolution graph embeddings, narrowing the performance gap with LLM. However, DRL shows minimal gains, as its policy network saturates in high-dimensional spaces, unable to exploit additional pilot resources effectively. Meanwhile, the LLM method dominates across all pilot lengths, attributed to its spatio-temporal attention mechanisms that adaptively prioritize critical RIMSA elements and users.  
The LLM framework excels by adaptively prioritizing critical RIMSA elements and users through its spatio-temporal attention mechanisms. This capability allows it to reach peak performance in data rates while requiring lower pilot resource consumption, thereby enabling efficient and effective large-scale RIMSA deployments without the need for excessive pilot lengths. Thus, even under conditions of reduced pilot overhead, the LLM approach manages to identify and implement the best possible RIMSA configurations, enhancing overall system performance in complex, multi-user environments.

\begin{figure}[t]
	\centering
	\includegraphics[width=3 in]{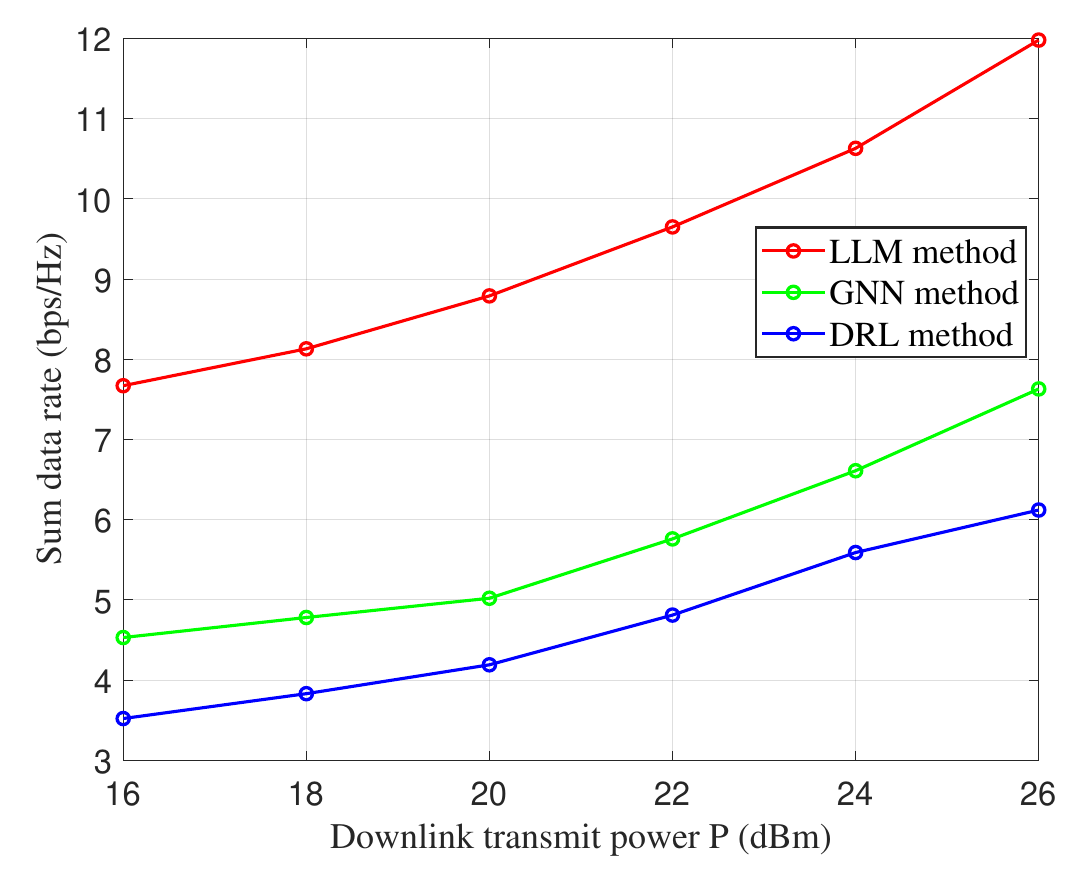}
	\caption{Performance of the proposed methods for RIMSA system with different transmitted power.}
	\label{sumrate_snr}
\end{figure}

The Fig. \ref{sumrate_snr} evaluates the sum data rate (bps/Hz) of three methods LLM, GNN, and DRL for a RIMSA system under varying downlink transmit power. The LLM method (red line) consistently achieves the highest performance, demonstrating its robustness in leveraging increased transmit power for adaptive beamforming and interference management. The GNN method (green line) follows as a sub-optimal solution, benefiting from its ability to model spatial interactions among RIMSA elements and users. In contrast, the DRL method (blue line) lags significantly, as its high-dimensional state-action space (due to multi-user, multi-RIMSA-element dynamics) hinders effective policy optimization. The LLM spatio-temporal attention mechanisms dynamically prioritize critical transmission paths, outperforming DRL and GNN, underscoring its adaptability to large scale metaelements control. GNN graph-structured learning captures spatial correlations, ensuring steady performance gains, but its reliance on fixed topology limits flexibility compared to LLM context-aware optimization. Meanwhile, DRL performance saturation reflects its inability to handle the combinatorial complexity of large-scale RIMSA systems, where state spaces grow exponentially with users and elements. These results highlight LLM dominance in high-power RIMSA deployments, while GNN remains a viable alternative for moderate scalability. The stark performance gaps emphasize the need for intelligent, topology-adaptive frameworks in next-generation networks.

\begin{figure}[t]
	\centering
	\includegraphics[width=3 in]{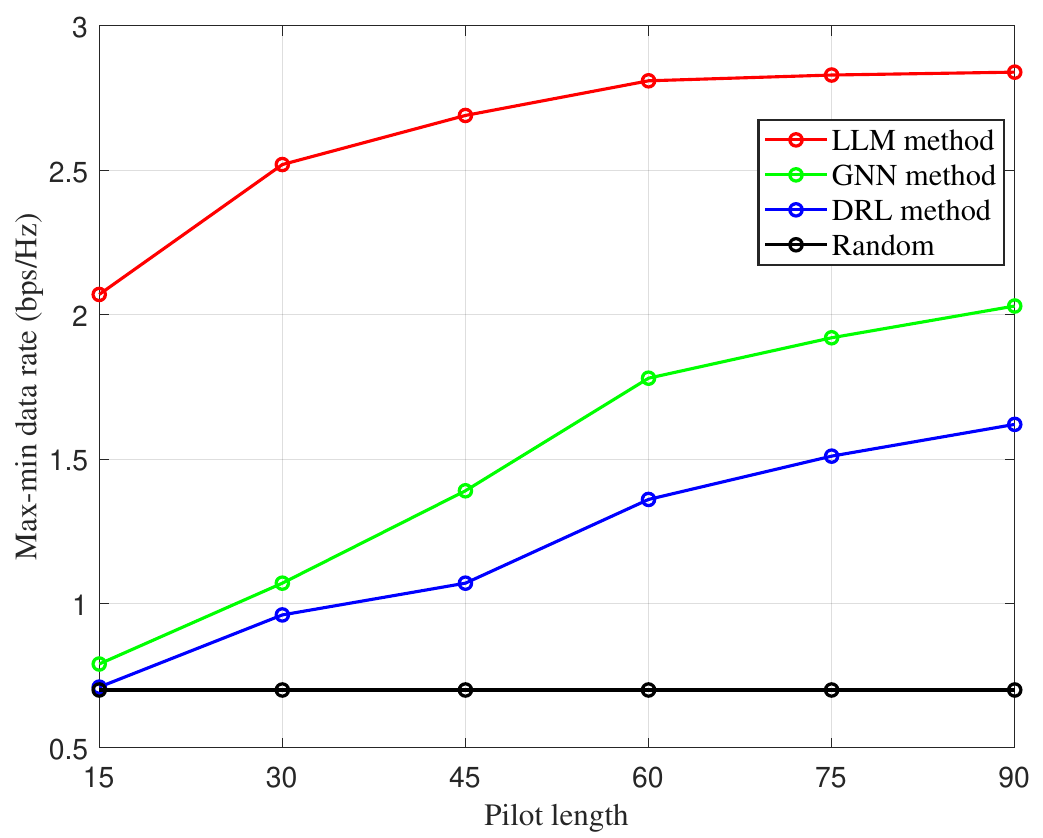}
	\caption{Performance of the proposed methods for RIMSA system with different pilot length.}
	\label{maxmin_pilot}
\end{figure}

The Fig. \ref{maxmin_pilot} evaluates the max-min data rate (bps/Hz) of four methods LLM, GNN, DRL, and random in a RIMSA system as pilot length increases, emphasizing how max-min optimization addresses communication fairness. The LLM method (red line) achieves the highest fairness performance, by explicitly optimizing the worst case user rate a critical metric for equitable service in multi-user networks. The GNN method (green line) follows with steady gains, while the DRL method (blue line) and random method (black line) lag significantly, underscoring their limitations in balancing fairness and efficiency. Max-min optimization is pivotal for fairness as it prioritizes the weakest user rate, ensuring no user is left behind due to interference or poor channel conditions;  This approach directly tackles fairness challenges in RIMSA systems, where users near cell edges or blocked by obstacles often suffer from severe rate imbalances. The LLM method excels by leveraging spatio-temporal attention mechanisms to dynamically allocate pilot resources and beamforming gains to underserved users, achieving performance growth and balancing spectral efficiency with fairness as more pilots become available. These results demonstrate that as highlighted by LLM dominance, showing how advanced algorithms can mitigate rate starvation for disadvantaged users and ensure uniform quality of service (QoS), a cornerstone for 6G networks demanding ultra-reliable and inclusive connectivity.

\begin{figure}[t]
	\centering
	\includegraphics[width=3 in]{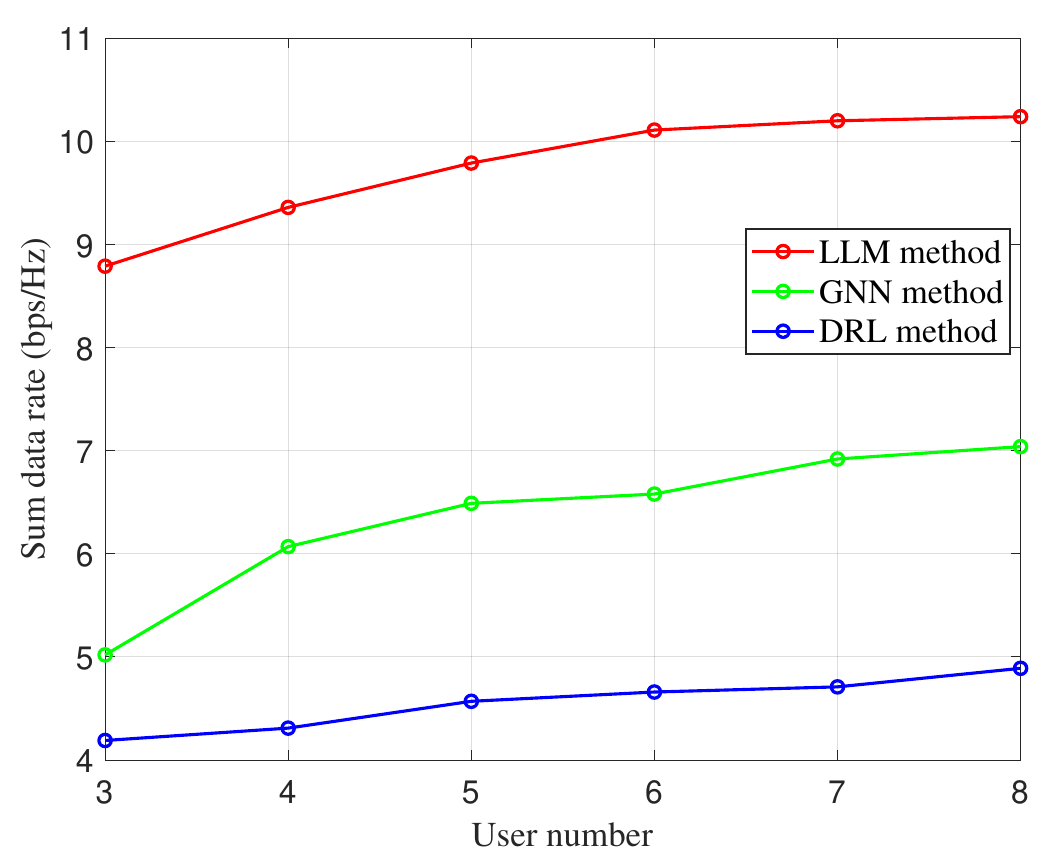}
	\caption{Performance of the proposed methods for RIMSA system with different user numbers.}
	\label{sumrate_user}
\end{figure}

The Fig. \ref{sumrate_user} evaluates the sum data rate (bps/Hz) of three methods in a RIMSA system with user numbers ranging from 3 to 8, where pilot length is explicitly set to 15$\times$ the user count (e.g., 45 for 3 users), reflecting the practical need to scale pilot resources proportionally with user density for sufficient channel estimation accuracy in multi-user environments. The LLM method (red line) maintains superior performance across all user scales, as its adaptive spatio-temporal attention dynamically optimizes pilot utilization while avoiding interference saturation despite growing pilot overhead. The GNN method (green line) shows steady growth by modeling user-RIMSA spatial correlations via graph learning; however, at higher user counts, its fixed graph topology struggles to fully exploit extended pilot sequences, limiting gains compared to LLM. In contrast, the DRL method (blue line) suffers from scalability bottlenecks, as the state-action space dimension grows exponentially with pilot length, rendering policy exploration intractable. The 15$\times$ pilot-user relationship highlights a fundamental trade-off: longer pilots enhance channel estimation but demand intelligent methods (like LLM) to mitigate overhead, while shorter pilots reduce overhead but risk inaccurate estimation, penalizing fairness and efficiency. The widening performance gaps between methods as users and pilots scale underscore LLM unique ability to balance spectral efficiency with pilot-resource constraints, making it indispensable for large scale RIMSA deployments.

\begin{figure}[t]
	\centering
	\includegraphics[width=3 in]{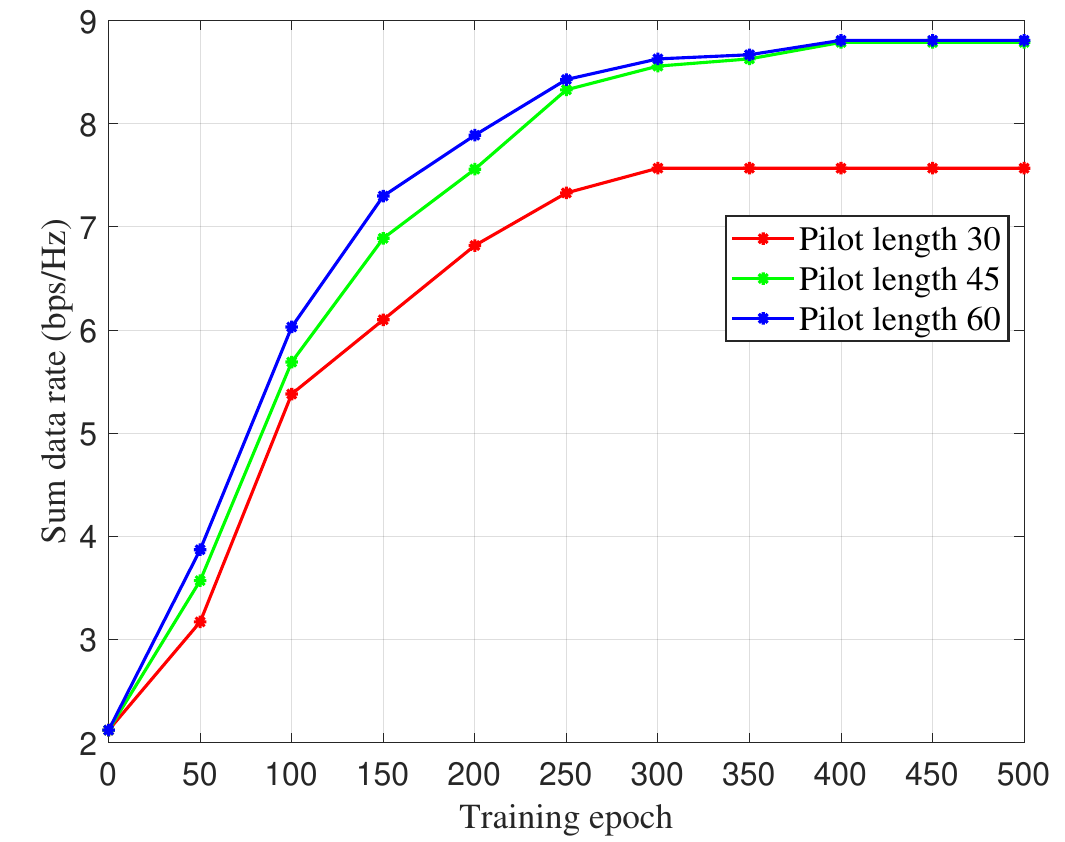}
	\caption{Performance of LLM for RIMSA system with different training epoch.}
	\label{sumrate_training}
\end{figure}
The Fig. \ref{sumrate_training} evaluates the sum data rate (bps/Hz) of the LLM  for the RIMSA  system across training epochs (0-500) under three pilot lengths: 30 (red), 45 (green), and 60 (blue). All curves exhibit a rapid initial improvement in data rate, gradually stabilizing after 250 epochs, reflecting the model convergence. Longer pilots consistently achieve higher performance, with the pilot length 60 (blue) curve outperforming pilot lengths 45 and 30, underscoring the critical role of extended pilots in enhancing channel estimation accuracy and interference management. Shorter pilots saturate earlier with limited gains, while longer pilots sustain gradual improvements until 350 epochs, leveraging additional training to optimize high-dimensional pilot-resource interactions. However, the extended training requirement for pilot length 45 highlights a trade-off between performance gains and computational complexity. These results validate that longer pilots significantly boost RIMSA system performance when paired with LLM adaptive training, but optimal deployment must balance spectral efficiency with training overhead.

\section{Conclusion}
This paper presents LLM-RIMSA, a new framework that integrates LLMs with RIMSA to address the limitations of conventional model-based optimization and DL in high-dimensional RIMSA control tasks. 
By resolving critical limitations of conventional RIS designs such as frequency selectivity, propagation delays, and restricted amplitude control, RIMSA enables unprecedented flexibility in dynamic beamforming and multi-user interference suppression. 
By leveraging the contextual reasoning and few-shot learning capabilities of LLMs, the framework eliminates dependency on explicit channel estimation while achieving real-time adaptive optimization of RIMSA configurations. Simulations demonstrate significant performance gains, low training and inference costs than traditional DL method.

This breakthrough transforms RIMSA from passive hardware into cognitive entities capable of self-optimization in complex radio environments. Future work will focus on LLM deployment for edge devices to advance LLM-driven intelligent surfaces toward practical 6G applications.





\end{document}